\documentclass{aastex}
\usepackage{spr-astr-addons}
\usepackage{url}
\usepackage{amssymb}

\RequirePackage{color}

\newcommand{\emaila}{hongwei.ge@gmail.com}

\begin{document}

\title{Stellar adiabatic mass loss model and applications}
\slugcomment{Accepted for publication in AP\&SS}
%% Running heads
\shorttitle{Adiabatic mass loss}
\shortauthors{Ge et al.}

\author{Hongwei Ge\altaffilmark{1,2,3}} \and \author{Ronald F. Webbink\altaffilmark{4}}
\and \author{Zhanwen Han\altaffilmark{1,2}}
%\affil{}
\and
\author{Xuefei Chen\altaffilmark{1,2}}
%\affil{}
%\email{\emaila}

\altaffiltext{1}{National Astronomical Observatories/Yunnan
Observatory, Chinese Academy of Sciences, Kunming, 650011, P.R. China}
\altaffiltext{2}{Key Laboratory for the Structure and Evolution of Celestial Objects, Chinese Academy of Sciences, Kunming 650011, P.R. China}
\altaffiltext{3}{Graduate University of Chinese Academy of Sciences, Beijing 100049, P.R. China}
\altaffiltext{4}{Department of Astronomy, University of Illinois, 1002 W. Green St. Urbana, IL 61801, U.S.A
\\E-Mail: \emaila}

\begin{abstract}
Roche-lobe overflow and common envelope evolution are very important in binary evolution, which is believed to be the main evolutionary channel to hot subdwarf stars. The details of these processes are difficult to model, but adiabatic expansion provides an excellent approximation to the structure of a donor star undergoing dynamical timescale mass transfer. We can use this model to study the responses of stars of various masses and evolutionary stages as potential donor stars, with the urgent goal of obtaining more accurate stability criteria for dynamical mass transfer in binary population synthesis studies. As examples, we describe here several models with the initial masses equal to $1~M_\odot$ and $10~M_\odot$, and identify potential limitations to the use of our results for giant branch stars.
\end{abstract}


%\section*{}
%\label{sec:intro}

\section{Introduction}
\label{sec1}

The Fourth Meeting on Hot Subdwarf Stars and Related Objects was convened in Shanghai, China from 19th to 24th, July. Recent discoveries and developments in both theory and observation of hot subdwarfs and related objects were reported, and many unsolved problems were discussed. As we all know, hot subdwarf stars are extreme horizontal-branch stars or related objects. They may dominate the UV-upturn of early-type galaxies and they exist in both the field of our Galaxy and its globular cluster system. About half of subdwarf B (sdB) stars are binaries \citep{max01,saf01}; binary evolution is obviously important in their formation, as the observed systems are too compact to have avoided past mass transfer. Binary population synthesis models \citep{han02,han03,han08} explain naturally the sdB binary fractions and the UV-upturn of early-type galaxies via: (a) one or two phases of common envelope (CE) evolution, (b) stable Roche-lobe overflow (RLOF), and (c) the merger of two He-core white dwarf stars (WDs).

RLOF and CE evolution are also very important in the formation of other binary systems, e.g., cataclysmic variables, X-ray binaries, double white dwarfs and binary neutron stars. Those systems containing compact objects are among the most energetic and rapidly variable sources known. Unfortunately, we know very little about the details of CE evolution, but dynamically unstable RLOF appears to be the trigger that launches it. The threshold conditions for dynamical mass transfer now in common use in binary population synthesis calculations are based on polytropic models for rapid mass loss process \citep[and references therein]{hjel87}. These polytropic studies provide useful qualitative insights into RLOF and CE evolution, but they omit much relevant physics, and fail to address many advanced evolutionary stages of interest. Detailed studies of binary evolution \citep{han02,chen08} reveal a need for more realistic determinations of the threshold conditions for dynamical mass transfer.

With the motivation and cautions above, we set out to study stellar rapid mass loss based on the pioneering work of \citet{hjel89a,hjel89b}. We describe the basic assumptions and numerical techniques employed in modeling stellar adiabatic mass loss in section~\ref{sec2}, with initial results and their possible application to binary population synthesis in section~\ref{sec3}. In section~\ref{sec4}, we offer a short discussion of some  remaining problems.

\section{Stellar adiabatic mass loss model}
\label{sec2}

Dynamical time scale mass transfer occurs when, in response to mass loss, the interior expansion of a donor star drives its surface beyond its Roche lobe.  That expansion is driven by the quasistatic readjustment of the star to the local drop in pressure as overlying mass is removed, and it is characterized by the star's dynamical time scale.  That time scale is typically many orders of magnitude shorter than either the nuclear or thermal time scale of the donor star.  In dynamical time scale mass transfer, therefore, the response of the donor star to mass loss becomes asymptotically adiabatic \citep{ge10,hjel87}, and its composition profile remains fixed.  Inside its Roche lobe, the donor remains in hydrostatic equilibrium, except very near the inner Lagrangian point. Indeed, even for thermal timescale mass transfer, thermal relaxation is largely confined to the outermost layers of the donor, and those layers are quickly stripped away. Model sequences in which potential donor stars are stripped of mass and allowed to respond only adiabatically therefore provide valuable insight into the inner structure of donor stars undergoing rapid mass transfer.  Furthermore, we can determine whether a given donor star will be stable or unstable to dynamical time scale mass transfer, by comparing its radius as it loses mass adiabatically to the effective radius of its Roche lobe.

The stellar structure equations for adiabatic mass loss models are time-independent. The equations for hydrostatic equilibrium and mass continuity are retained, but those for energy conservation and energy transport are replaced by algebraic constraints fixing the entropy and composition profiles:
\begin{equation}
\frac{\partial \ln P}{\partial m}= - \frac{G m}{4 \pi r^4 P},
\label{pressure}
\end{equation}
\begin{equation}
\frac{\partial r^2}{\partial m}= \frac{1}{2 \pi r \rho},
\label{radius}
\end{equation}
\begin{equation}
s(m) = s_0(m),
\label{entropy}
\end{equation}
\begin{equation}
X(m) = X_0(m),
\label{composition}
\end{equation}
with structure variables $P$ for pressure, $r$ for radius, $m$ for mass, $T$ for temperature, $s$ for specific entropy and $X$ for the abundances of various nuclear species. The initial specific entropy $s_0$, and abundances $X_0$ are tabulated as functions of the radial mass coordinate $m$ in the initial model for an adiabatic mass loss sequence, and remain unchanged.

The solution to these equations not only describes the dynamical response of a star to mass loss, but it provides valuable insights into how that star then relaxes thermally in response to adiabatic expansion.  Because the specific entropy and composition are fixed as functions of mass, so too are their gradients with respect to mass.  One can write the local energy flux through the stellar interior in terms of the local entropy gradient, and thereby reconstruct the interior luminosity profile.  Gradients in that luminosity profile in turn imply a specific disposition of energy sources and sinks through the stellar interior.

\section{Initial results}
\label{sec3}

Based on the stellar evolution code developed by \citet{egg71,egg72,egg73} and \citet{pax04}, we built a numerical code in FORTRAN95 to calculate stellar adiabatic mass loss models. The latest input physics of our code are the same as in Eggleton's code \citep{han94,han03,pols95,pols98}. We set the mixing-length parameter of $\alpha = l/H_{\rm P} = 2.0$ and the convective overshooting parameter $\delta_{\rm ov} = 0.12$ \citep{pols98}.  With our stellar adiabatic mass loss code, we plan to study the responses of donor stars to mass loss at different evolutionary stages and metallicities of interest.  As an initial application, we describe here two sets of models with metallicity $Z=0.02$, initial masses $1~M_\odot$ and $10~M_\odot$, respectively.

\subsection{$1~M_\odot$ star: radiative core with a convective envelope}

\begin{figure}[t]
\scalebox{0.45}[0.45]{ \rotatebox{0}{\includegraphics{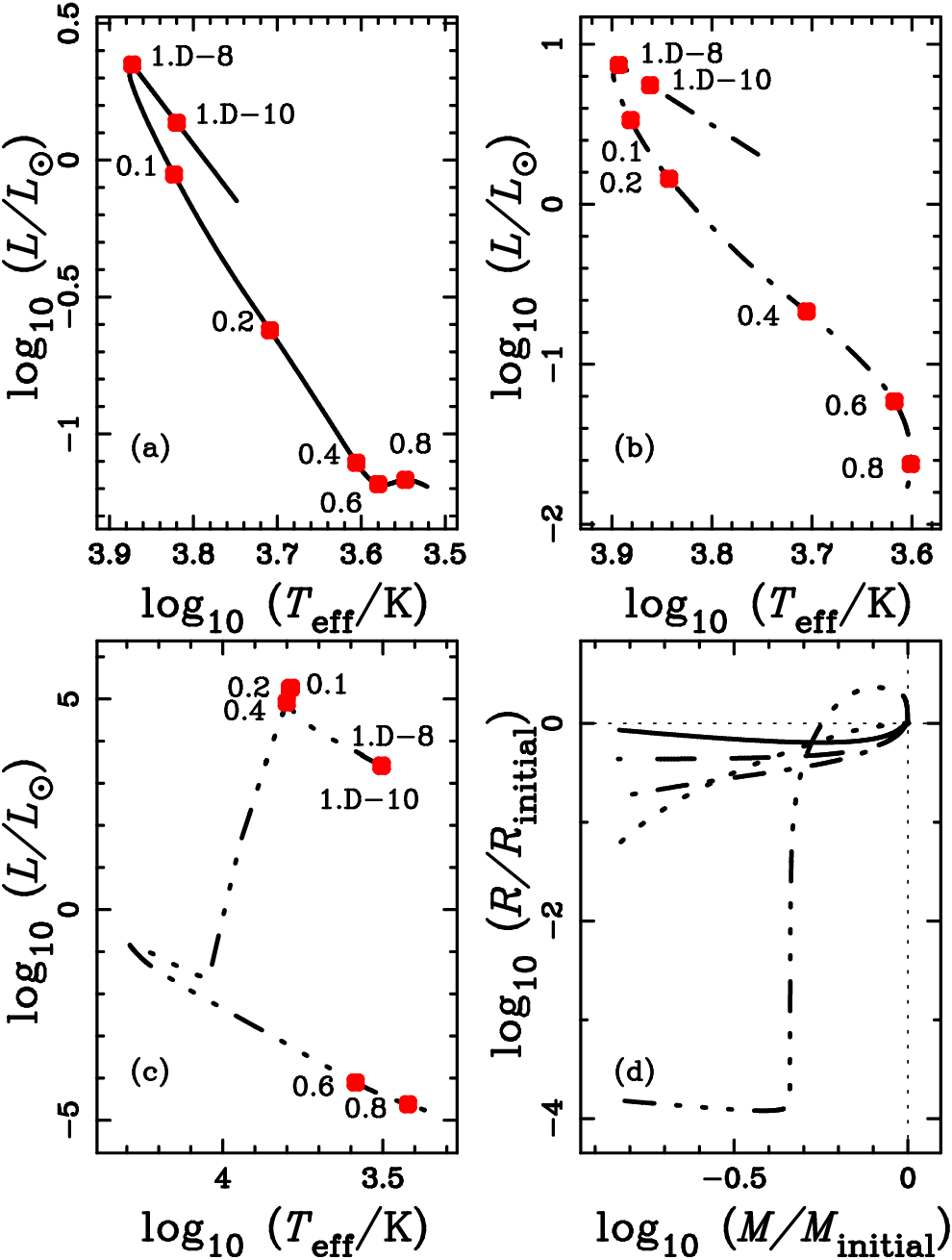}}}
\caption{Hertzsprung-Russell (HR) diagrams and radial responses of $1~M_\odot$ models during adiabatic mass loss. Frames (a), (b) and (c) show the tracks in the HR diagram of stars initiating mass loss at the zero-age main sequence (ZAMS -- solid line), in the Hertzsprung gap (HG -- dash-dot line), and at the tip of the red giant branch (TGB -- dash-dot-dot-dot line), respectively. The solid squares and numbers in each of these frames indicate the amount of mass lost (in $M_\odot$) at that point in the track. Frame (d) shows the radial responses to mass loss for each of the models shown in frames (a)-(c), with the addition of models at the terminal main sequence (TMS -- dashed line) and at the base of the red giant branch (RGB -- dotted line). Models with relatively shallow surface convection zones (ZAMS, TMS, and HG) initially contract in response to mass loss, while those with deep convective envelopes (RGB and TGB) expand, dramatically so when convection becomes inefficient and the outer layers have strongly superadiabatic temperature gradients.}
\label{fig1}
\end{figure}

\begin{figure}[t]
\scalebox{0.37}[0.37]{ \rotatebox{-90}{\includegraphics{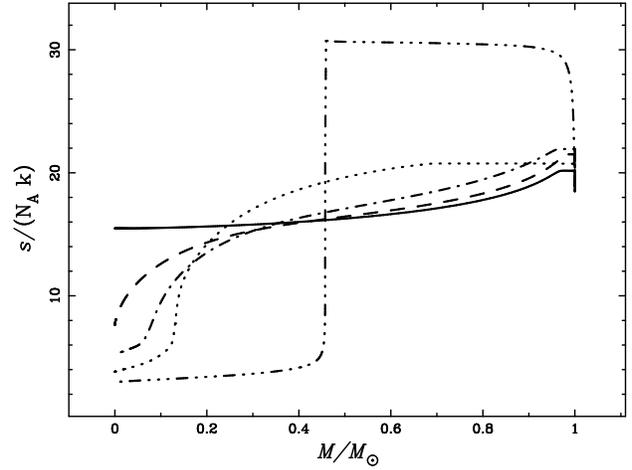}}}
\caption{The specific entropy profiles of the initial $1~M_\odot$ models of the mass-loss sequences shown in Fig.~\ref{fig1}.  Lines are coded as in Fig.~\ref{fig1} according to evolutionary phase at the beginning of each mass loss sequence. As the star evolves, the specific entropy in its core decreases.  A sharp positive entropy gradient develops just outside the core with the onset of hydrogen shell burning, becoming more pronounced with time.  The steep negative entropy gradients at the surfaces of more evolved stars are due to superadiabatic (inefficient) convection.}
\label{fig2}
\end{figure}

In its simplest terms, the adiabatic response of a star to mass loss depends on the star's specific entropy profile.  In radiatively stable zones, entropy increases outwards, and the removal of high-entropy surface layers brings lower-entropy (denser) material toward the surface, leading to contraction; in convective zones, the entropy profile is nearly flat (except possibly for superadiabatic surface layers) and the star expands in response to mass loss.
Stars typically consist of both radiative and convective layers, so there is a competition between these effects, but one should note that, in strictly adiabatic mass loss, the entropy profile remains fixed in mass, and so convective boundaries also remain fixed in mass.
With these concepts in mind, let us examine the behaviour of a 1~$M_\odot$ star in different evolutionary stages as it loses mass.  Such a star has a surface convection zone throughout its lifetime, but that zone contains relatively little mass on the main sequence, deepening dramatically as the star reaches the giant branch.  Fig.~\ref{fig1} shows the response of a 1~$M_\odot$ star to mass loss at different evolutionary stages:

\begin{enumerate}

\item Surface convection zones are invariably capped by a superadiabatic layer in which convection becomes very inefficient.  Until the star begins ascending the giant branch, however, the mass in this superadiabatic zone is quite small (about $10^{-8}\ M_\odot$); the star brightens as higher-entropy material is exposed, but its change in radius is negligible (see (a), and (b) in Fig.~\ref{fig1}).

\item ZAMS: Chemically-homogeneous lower main sequence stars are reasonably well-described by a composite polytrope model \citep{rap83,hjel87}. This $1~M_\odot$ ZAMS star contracts because of the sharp specific entropy gradient in the outer part of its radiative core. After about half of its initial mass is lost, however, the entropy profile of the remaining star has become very flat (see solid line in Fig.~\ref{fig2}), and the star begins to expand in response to mass loss.

\item ZAMS to TMS: As the star evolves through the main sequence, the specific entropy of the core decreases, while that of the outer envelope increases (see dashed line in Fig.~\ref{fig2}), and the contraction in response to mass loss becomes more pronounced.

\item TMS to RGB: The mass and depth of the convective envelope increases dramatically as the star reaches the RGB. Its radial response becomes dominated by its convective envelope, and star expands modestly in response to mass loss.

\item RGB to TGB: Though the mass fraction of the convective envelope becomes smaller, its specific entropy grows enormously as the star evolves; convection becomes very inefficient near its surface, and the star develops an extended superadiabatic surface layer. That layer dominates the initial response of the star to mass loss, leading to a rapid initial expansion. The response of the stellar radius to mass loss does not show a simple relation with mass fraction of the convective envelope as do main sequence stars.

\end{enumerate}

\subsection{$10~M_\odot$ star: convective core with a radiative envelope}

\begin{figure}[t]
\scalebox{0.45}[0.45]{ \rotatebox{0}{\includegraphics{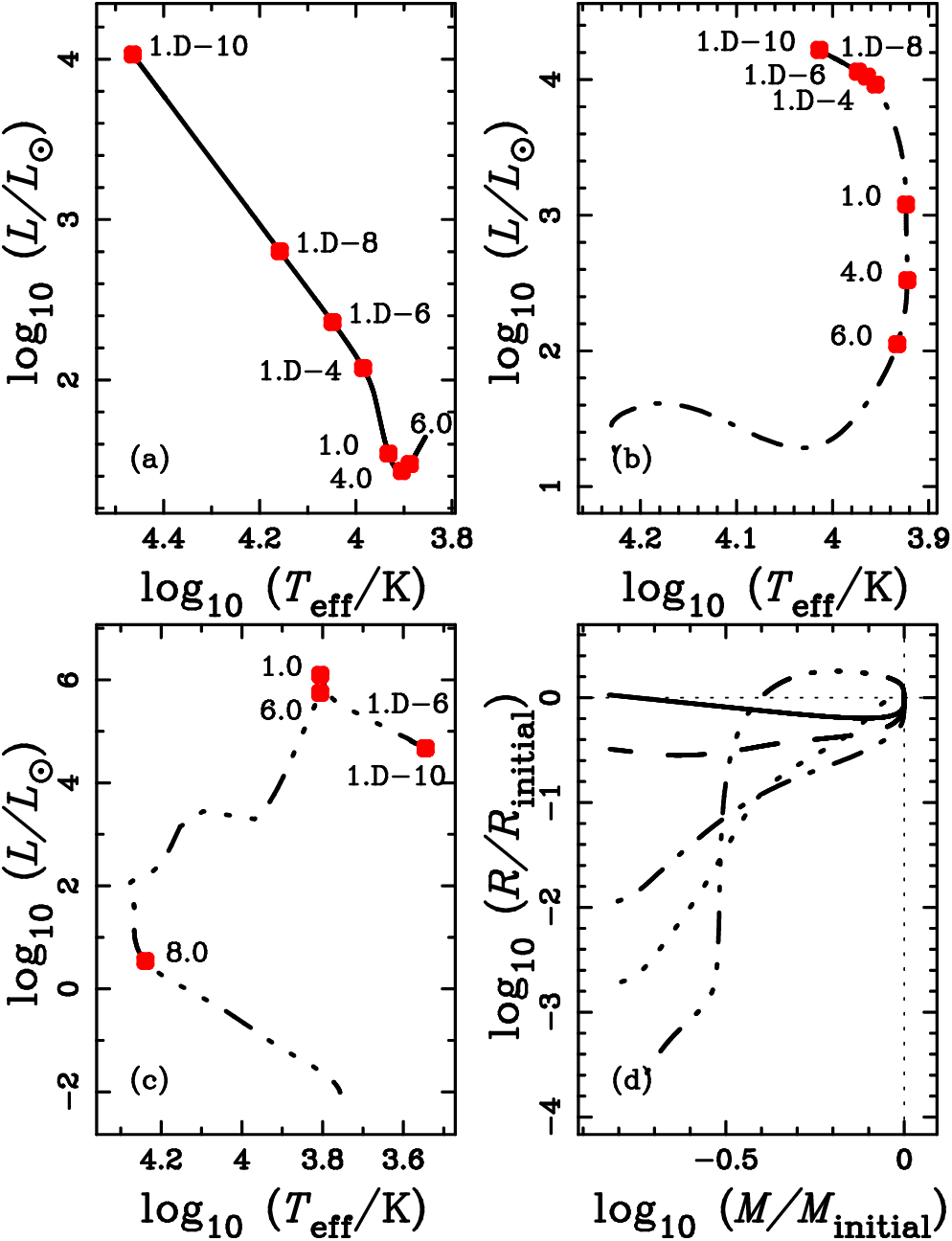}}}
\caption{HR diagrams and the radial responses of $10~M_\odot$ models during adiabatic mass loss. Frames (a), (b), and (c) show tracks in the HR diagram of stars initiating mass loss at ZAMS, HG, and TGB phases, respectively, as in Fig.~\ref{fig1}. The solid squares and numbers in each of these frames indicate the amount of mass lost (in $M_\odot$) at that point in the track. Frame (d) shows the radial responses to mass loss for each of the models shown in frames (a)-(c), as well as for TMS and RGB models. Lines are coded according to evolutionary phase as in Fig.~\ref{fig1}.}
\label{fig3}
\end{figure}

\begin{figure}[t]
\scalebox{0.37}[0.37]{ \rotatebox{-90}{\includegraphics{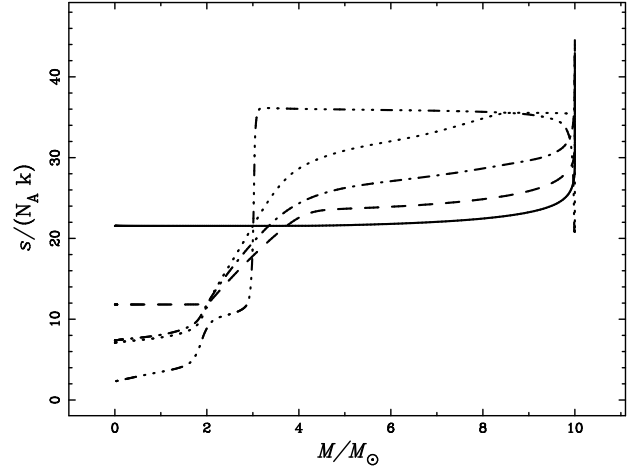}}}
\caption{The specific entropy profiles of the initial $10~M_\odot$ models of the mass-loss sequences shown in Fig.~\ref{fig3}. Lines are coded as in Fig.~\ref{fig1} according to evolutionary phase at the beginning of each mass loss sequence. As seen in Fig.~\ref{fig2}, the specific entropy in the core decreases as the central composition changes. The flat entropy profiles in the core are due to convection, while the steep positive gradients at the surface are due to the radiative envelope. As seen in Fig.~\ref{fig2}, sharp positive entropy gradients are seen outside hydrogen-burning shells, and in the case of the TGB model, we see also the sharp increase outside the helium-burning shell.  The TGB model again has a very extended, superadiabatic envelope, reflected in the sharp negative entropy gradients near the surface.}
\label{fig4}
\end{figure}

Although the $10~M_\odot$ models have very different inner structure from the $1~M_\odot$ models, the behaviour of these models (see Fig.~\ref{fig3}) can be understood with the same general precepts:

\begin{enumerate}

\item The relatively massive $10~M_\odot$ stars lack a surface convection zone until late in their evolution.  Specific entropy increases dramatically very near the surface, as seen in Fig~\ref{fig4}. As these layers are stripped away, the star decreases rapidly in effective temperature and luminosity as lower-entropy material is exposed, but the amount of mass involved is still small (less than about $10^{-4}~M_{\odot}$), and so the stellar radius changes very little at first.

\item ZAMS to TMS: $10~M_\odot$ main-sequence stars have pronounced positive entropy gradients in their radiative envelopes, leading to stronger initial contraction than seen above in the $1~M_\odot$ models.  However, they also have large convective cores, with flat entropy profiles that ensure they re-expand with sufficient mass loss, like the $1~M_\odot$ models.

\item TMS to RGB: As $10~M_\odot$ stars evolve across the HR diagram, the rapid rise in specific entropy through the envelope softens.  Their initial contraction in response to mass loss becomes less severe, but the growing entropy difference between core and envelope (see Fig.~\ref{fig4}) suppresses re-expansion.

\item RGB to TGB: Like $1~M_\odot$ stars, $10~M_\odot$ stars with deep convective envelopes expand in response to mass loss (compare Fig.~\ref{fig3} (d) and Fig.~\ref{fig4} (d)).  At their high luminosities and large radii, they, too, develop superadiabatic surface layers, and expand violently in response to its removal.

\end{enumerate}

\subsection{Application to binary population synthesis}

\begin{figure}[t]
\scalebox{0.37}[0.37]{ \rotatebox{-90}{\includegraphics{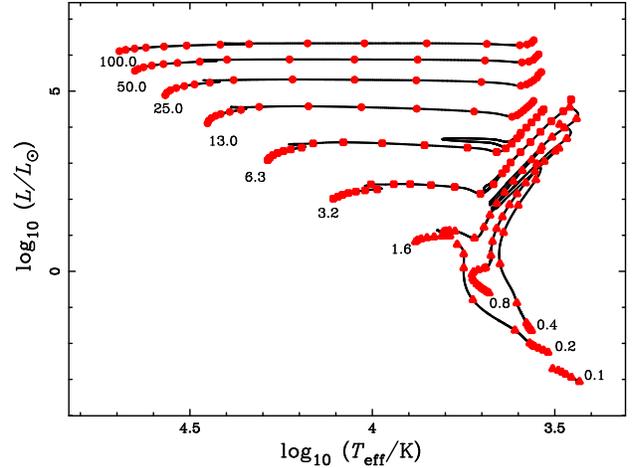}}}
\caption{The model grid for different stars with metallicity $Z=0.02$. The full set of models (not shown) ranges in mass from $0.1~M_\odot$ to $100~M_\odot$, at intervals of $\log M = 0.1$. Solid circles, squares and triangles indicate different evolution stages for high, intermediate and low mass stars, respectively. They mark initial models for adiabatic mass loss sequences.}
\label{fig5}
\end{figure}

When a star fills its Roche lobe ($R = R_L$), the stability of the binary against dynamical time scale mass transfer depends on whether adiabatic perturbations to the system are damped or unstable.  That criterion for stability can be written as  \citep{webb85}
\begin{equation}
\zeta_{\rm ad} > \zeta_L,
\label{pressure}
\end{equation}
where $\zeta_{\rm ad} \equiv (\partial \ln R/\partial \ln M)_{\rm ad}$ is the stellar adiabatic radius-mass exponent, and $\zeta_L \equiv (\partial \ln R_L/\partial \ln M)_{\rm ad}$ is the time-independent first-order response of the Roche lobe to mass transfer. In the simplest models of mass transfer, in which systemic mass and orbital angular momentum loss rates are proportional to the mass transfer rate, $\zeta_L$ depends only on the binary mass ratio. Our stellar adiabatic mass loss models therefore allow us to determine the critical mass ratio (donor/accretor) above which a binary will be unstable to dynamical time scale mass transfer. This is an essential consideration in mapping out binary population synthesis models. We are currently building a library of stellar adiabatic mass loss models \citep{ge10}, covering the full range of evolutionary stages from ZAMS to the tip of the RGB or AGB, as appropriate. Fig.~\ref{fig5} shows the model grid of our project.

\section{Discussion and conclusions}
\label{sec4}
RLOF process and CE evolution are very important in the formation of hot subdwarf stars and other binary systems. Earlier published studies of \citet[and references therein]{hjel87} provided useful qualitative insights in RLOF and CE evolution, but left considerable room for improvement, especially with regard to the range of evolved phases that need to be addressed. These deficiencies were in part redressed in later work \citep{hjel89b}, but only a fragment of that work was ever published \citep{hjel89a}.  The stellar adiabatic mass loss models described in this paper considerably extend the scope of Hjellming's work.  They allow us not only to study the interior structure of donor stars undergoing dynamical timescale mass transfer, but also to evaluate the stability criteria for dynamical mass transfer in binary population synthesis.

An initial application of our models to $1~M_\odot$ and $10~M_\odot$ stars has been described in this paper.  The reader should beware that the these models have limitations to their usefulness, as they depend on being able to separate the process of dynamical relaxation to hydrostatic equilibrium from that of thermal relaxation to thermal equilibrium.  On the main sequence, for example, the dynamical time scale of the Sun is roughly $10^{-10}$ that of its thermal time scale, making the adiabatic mass loss model an excellent description of the asymptotic behavior of a solar-type star to very rapid mass loss.  However, at the opposite extreme in the HR diagram, stars near the tip of the RGB or AGB have thermal time scales approaching their dynamical time scales, and the entire mass transfer process may require full-blown time-dependent modeling. This convergence of thermal and dynamical time scales is broadly related to the abrupt expansion seen in our models of $1~M_\odot$ and $10~M_\odot$ stars at the tip of the giant branch (TGB), when their outermost superadiabatic layers are stripped away.  In reality, Roche lobe overflow is far from spherically-symmetric, as treated here (by necessity), but that superadiabatic expansion may nevertheless reflect a real physical phenomenon.  Convection itself is not spherically symmetric, and with convective velocities in the superadiabatic zone approaching sound speed, it may be possible for rising flows near the inner Lagrangian point to bridge the potential to the companion's Roche lobe while the donor still lies well within its own Roche limit.  We are not now able to pass judgement on that possibility.

When dynamical instability occurs, common envelope evolution almost certainly follows, as the thermal time scale of the accreting star is invariably longer than that of the donor, which itself is generally much longer than its dynamical time scale.  This ordering of time scales ensures that the envelope cannot cool efficiently on the transfer time scale, but remains extended and engulfs both stellar cores.  It may happen that thermal time scale mass loss from the donor is still rapid enough to form a common envelope, but this case can only arise if the system is stable against dynamical mass transfer, while unstable to thermal time scale mass transfer. Since the stellar dynamical time scales of both donor and accretor are shorter than their thermal time scales, the prospect arises of a quasistatic common envelope, that is, of formation of a contact binary.  Such large numbers of such objects are known (the W UMa systems, as examples) that they must be very long-lived, evolving in a very different fashion from CE evolution as we have used that term above.

We believe that stellar adiabatic mass loss models provide the most useful approach to date toward defining the limits of dynamical stability in interacting binaries, essential input to the construction of binary population synthesis models.

\acknowledgments
This work is supported by the National Natural Science Foundation of China (grant Nos. 10821061, 10973036 and 2007CB815406), the U.S. National Science Foundation (grant AST 0406726), the Chinese Academy of Sciences under Grant No. KJCX2-YW-T24, and the Yunnan Natural Science Foundation (Grant No. 08YJ041001). We thank Philipp Podsiadlowski for insightful comments. We also thank the anonymous referee for his valuable comments that helped us to improve the paper.

\end{document}